\documentclass[conference,a4paper,fleqn]{IEEEtran}
\usepackage[cmex10]{amsmath}
\usepackage{amssymb}
\usepackage{hyperref}
\hypersetup{%
  pdftitle={Marginalizing over the PageRank Damping Factor},
  pdfauthor={Christian Bauckhage},
  pdfsubject={information retrieval, classification},
  pdfkeywords={PageRank, TotalRank, matrix logarithm},
  colorlinks=true,
  urlcolor=blue,
  citecolor=green,
  bookmarks=false}
  
\newcommand{\mat}{\boldsymbol}
\renewcommand{\vec}{\boldsymbol}

\begin{document}

\title{Marginalizing over the PageRank Damping Factor}

\author{%
\IEEEauthorblockN{Christian Bauckhage}
\IEEEauthorblockA{B-IT, University of Bonn, Germany \\
    Fraunhofer IAIS, Sankt Augustin, Germany}}

\maketitle

\begin{abstract}
In this note, we show how to marginalize over the damping parameter of the PageRank equation so as to obtain a parameter-free version known as TotalRank. Our discussion is meant as a reference and intended to provide a guided tour towards an interesting result that has applications in  information retrieval and classification.
\end{abstract}

\section{Introduction}

The \textit{PageRank} algorithm \cite{Page1999-TPR} is of fundamental importance in Web search \cite{Brin1998-TAO,Langville2006-GPR} and (multi-media) information retrieval \cite{Manning2008-ITI,Bauckhage2007-DFI,Jing2008-VAP} and provides an approach to problems in pattern recognition \cite{Bauckhage2008-PDC} or probabilistic inference \cite{Neumann2011-MLS}. Here, we show how to derive \textit{TotalRank}, a variant of PageRank that was first proposed in \cite{Boldi2005-TRR} and independently discussed in \cite{Bauckhage2008-PDC}. For brevity, we assume that the reader has a working knowledge of discrete time Markov chains and the basic ideas behind PageRank. A thorough and very readable introduction to these topics can, for instance, be found in the excellent book by Langville and Meyer \cite{Langville2006-GPR}.

\section{PageRank Equations}

Recall that the PageRank paradigm is concerned with the long-term behavior of the following dynamic process
\begin{equation}
    \vec{o}_{t+1} = \alpha \mat{H} \, \vec{o}_t + (1-\alpha) \vec{o}_0 \label{eq:iterative}
\end{equation}
where each of the vectors $\vec{o} \in \mathbb{R}^n$ is a stochastic vector, the matrix $\mat{H} \neq \mat{I} \in \mathbb{R}^{n \times n}$ is a Markov matrix, and the scalar $\alpha \in \mathbb{R}$ obeys $0 < \alpha < 1$. Unrolling the recursion in \eqref{eq:iterative}, we find the closed form expression
\begin{align}
\label{eq:closed}
%\vec{o}_1 & = \alpha \mat{H} \, \vec{o}_0 + (1-\alpha) \vec{o}_0 \notag \\
%\vec{o}_2 & = \alpha \mat{H} \, \bigl( \alpha \mat{H} \, \vec{o}_0 + (1-\alpha) \vec{o}_0 \bigr) + (1-\alpha) \vec{o}_0 \notag \\
%          & = \bigl[ \alpha \mat{H} \bigr]^2 \, \vec{o}_0 + (1-\alpha) \sum_{i=0}^{1} \bigl[ \alpha \mat{H} \bigr]^i \vec{o}_0 \notag \\
%          & \;\; \vdots \notag \\
\vec{o}_t & = \bigl[ \alpha \mat{H} \bigr]^t \vec{o}_0 + (1-\alpha) \sum_{i = 0}^{t-1} \bigl[ \alpha \mat{H} \bigr]^i \vec{o}_0         
\end{align}
and ask for the limit $\vec{o}_\infty = \lim_{t \rightarrow \infty} \vec{o}_t$ of this process. 

Since $\mat{H}$ is stochastic, its spectral radius $\rho(\mat{H}) = 1$. If it is also \textit{irreducible} and \textit{primitive} (e.g. has at least one positive diagonal element), the limit $\mat{H}_{\infty} = \lim_{t \rightarrow \infty} \mat{H}^t$ exists. In this case, we have
\begin{equation}
\label{eq:final1}
\lim_{t \rightarrow \infty} \bigl[ \alpha \mat{H} \bigr]^t = \mat{0}
\end{equation}
as well as
\begin{equation}
\label{eq:final2}
\lim_{t \rightarrow \infty} \sum_{i=0}^{t-1} \bigl[ \alpha \mat{H} \bigr]^i = \Bigl[ \mat{I} - \alpha\mat{H} \Bigr]^{-1}
\end{equation}
and therefore find convergence to
\begin{equation}
  \label{eq:final}
  \vec{o}_\infty = (1-\alpha) \Bigl [ \mat{I} - \alpha\mat{H} \Bigr ]^{-1} \vec{o}_0. 
\end{equation}

Looking at \eqref{eq:final}, we recognize that the \textit{PageRank vector} $\vec{o}_\infty$ depends on the \textit{damping factor} $\alpha$ which allows for trading off effects due to the \textit{transition matrix} $\mat{H}$ and the \textit{personalization} or \textit{teleportation vector} $\vec{o}_0$. Different choices of $\alpha$ can therefore lead to significantly different results \cite{Bressan2010-CTD} and while the problem of how to choose $\alpha$ in practical applications has been studied extensively \cite{Bressan2010-CTD,BaezaYates2006-GPR,Gleich2014-ADS}, a definitive answer remains elusive. Hence, an interesting alternative is to try to avoid choosing $\alpha$ altogether, for example by means of averaging. This idea was first considered in \cite{Boldi2005-TRR} (where it was termed \textit{TotalRank}), yet, neither \cite{Boldi2005-TRR} nor 
\cite{Bauckhage2008-PDC} (where it was applied in practice) provide details as to how to derive the resulting equation. While the required algebraic manipulations of \eqref{eq:final} are not overly complicated they are not exactly textbook material either. Below, we therefore discuss in detail how to obtain the TotalRank equation.

\section{Step-by-Step Derivation of TotalRank}

Our goal in this section is to eliminate the damping factor $\alpha$ from \eqref{eq:final}. One way of achieving this consists in marginalizing over $\alpha$ and requires us to evaluate the definite integral
\begin{equation}
  \int_0^1 \vec{o}_\infty \; d\alpha = \int_0^1 (1-\alpha) \Bigl [ \mat{I} - \alpha\mat{H} \Bigr ]^{-1} \vec{o}_0 \; d\alpha.
\end{equation}

Using \eqref{eq:final2}, we find that this apparently daunting integral which involves an inverted matrix can be written as
\begin{align}
\int_0^1 \vec{o}_\infty \; d\alpha
& = \int_0^1 (1-\alpha) \sum_{t=0}^\infty \alpha^t \mat{H}^t \, \vec{o}_0 \; d\alpha \nonumber \\ %\label{eq:deriv1} \\
& = \sum_{t=0}^\infty \left( \int_0^1 (1-\alpha) \alpha^t \; d\alpha \right) \mat{H}^t \, \vec{o}_0 \label{eq:deriv2}
\end{align}
and note that the integral which appears inside of the infinite series in \eqref{eq:deriv2} is rather elementary and evaluates to
\begin{equation}
\int_0^1 (1-\alpha) \alpha^t \; d\alpha  = \Bigl (  \frac{1}{t+1} - \frac{1}{t+2}\Bigr ).
\end{equation}

Plugging this result back into \eqref{eq:deriv2} yields
\begin{equation}
  \label{eq:deriv3}
  \int_0^1 \vec{o}_\infty \; d\alpha = \sum_{t=0}^\infty \Bigl (  \frac{1}{t+1} - \frac{1}{t+2}\Bigr ) \mat{H}^t \, \vec{o}_0
\end{equation}
which is indeed an expression in which the damping parameter $\alpha$ does not appear anymore. However, as the right hand side of \eqref{eq:deriv3} consists of an infinite matrix series, it seems of limited practical use because it is not immediately clear how to implement it on a computer.

We therefore continue with our efforts and consider the two matrix series 
\begin{align}
\sum_{t=0}^\infty \frac{\mat{H}^t}{t+1} 
& = \mat{H}^0 + \sum_{t=1}^\infty \frac{\mat{H}^t}{t+1} \nonumber \\
& = \mat{I} + \mat{H}^{-1} \sum_{t=1}^\infty \frac{\mat{H}^{t+1}}{t+1} \label{eq:deriv4}
\intertext{and}
\sum_{t=0}^\infty \frac{\mat{H}^t}{t+2}
& = \frac{1}{2} \mat{H}^0 + \sum_{t=1}^\infty \frac{\mat{H}^t}{t+2} \nonumber \\
& = \frac{1}{2} \mat{I} + \mat{H}^{-2} \sum_{t=1}^\infty \frac{\mat{H}^{t+2}}{t+2} \label{eq:deriv5}
\end{align}
where, in \eqref{eq:deriv5}, we use the notation $\mat{H}^{-2}$ to indicate the matrix product $\mat{H}^{-1} \mat{H}^{-1}$.

Given these forms of the series that appear in \eqref{eq:deriv3}, we recall the following representation of the \textit{matrix logarithm}
\begin{equation}
\log (\mat{I} - \mat{H}) = - \sum_{t=1}^\infty \frac{\mat{H}^t}{t}
\end{equation}
which is well defined, if $\rho(\mat{H})<1$. The expression in \eqref{eq:deriv4} can thus be written as
\begin{align}
  & \mat{I} + \mat{H}^{-1} \Bigl [ -\log(\mat{I}-\mat{H}) - \mat{H}\Bigr ] \nonumber \\
  & = \mat{I} - \mat{H}^{-1} \log(\mat{I}-\mat{H}) - \mat{H}^{-1} \mat{H} \nonumber \\
  & = - \mat{H}^{-1} \log(\mat{I}-\mat{H})
\end{align}
and \eqref{eq:deriv5} becomes
\begin{align}
& \frac{1}{2} \mat{I} + \mat{H}^{-2} \Bigl [ -\log(\mat{I}-\mat{H}) - \mat{H} -\frac{1}{2} \mat{H}^2 \Bigr ] \nonumber \\
& = \frac{1}{2} \mat{I} - \mat{H}^{-2} \log(\mat{I}-\mat{H}) - \mat{H}^{-2} \mat{H} - \frac{1}{2} \mat{H}^{-2} \mat{H}^2 \nonumber \\
& = - \mat{H}^{-2} \log(\mat{I}-\mat{H}) - \mat{H}^{-1}.
\end{align}
Hence, if we subtract these two expressions as required by \eqref{eq:deriv3}, we obtain
\begin{align}
& \mat{H}^{-2} \log(\mat{I}-\mat{H}) + \mat{H}^{-1} - \mat{H}^{-1} \log(\mat{I}-\mat{H}) \nonumber \\
& = \mat{H}^{-1} \Bigl [ \mat{I} + \mat{H}^{-1} \log(\mat{I}-\mat{H}) - \log(\mat{I}-\mat{H}) \Bigr ] \nonumber \\
& = \mat{H}^{-1} \Bigl [ \mat{I} + \bigl[ \mat{H}^{-1} - \mat{I} \bigr] \log(\mat{I}-\mat{H}) \Bigr ].
\end{align}

This establishes our final result: averaging away the damping factor in the PageRank equation \eqref{eq:final} leads to the following clsoed form solution 
\begin{equation}
\int_0^1 \vec{o}_\infty \; d\alpha = \mat{H}^{-1} \Bigl [ \mat{I} + \bigl[ \mat{H}^{-1} - \mat{I} \bigr] \log(\mat{I}-\mat{H}) \Bigr ] \vec{o}_0.
\end{equation}

\section{Conclusion}

In this brief note, we demonstrated how to marginalize over the damping parameter $\alpha$ in the PageRank equation. Several tedious yet straightforward algebraic manipulations led to a pleasantly simple closed form solution that involves a matrix logarithm.

\bibliographystyle{IEEEtran}
\bibliography{literature}

% Generated by IEEEtran.bst, version: 1.13 (2008/09/30)
\begin{thebibliography}{10}
\providecommand{\url}[1]{#1}
\csname url@samestyle\endcsname
\providecommand{\newblock}{\relax}
\providecommand{\bibinfo}[2]{#2}
\providecommand{\BIBentrySTDinterwordspacing}{\spaceskip=0pt\relax}
\providecommand{\BIBentryALTinterwordstretchfactor}{4}
\providecommand{\BIBentryALTinterwordspacing}{\spaceskip=\fontdimen2\font plus
\BIBentryALTinterwordstretchfactor\fontdimen3\font minus
  \fontdimen4\font\relax}
\providecommand{\BIBforeignlanguage}[2]{{%
\expandafter\ifx\csname l@#1\endcsname\relax
\typeout{** WARNING: IEEEtran.bst: No hyphenation pattern has been}%
\typeout{** loaded for the language `#1'. Using the pattern for}%
\typeout{** the default language instead.}%
\else
\language=\csname l@#1\endcsname
\fi
#2}}
\providecommand{\BIBdecl}{\relax}
\BIBdecl

\bibitem{Page1999-TPR}
L.~Page, S.~Brin, R.~Motwani, and T.~Winograd, ``{The PageRank Citation
  Ranking: Bringing Order to the Web},'' Stanford InfoLab, Tech. Rep. 422,
  1999.

\bibitem{Brin1998-TAO}
S.~Brin and L.~Page, ``{The Anatomy of a Large-scale Hypertextual Web Search
  Engine},'' \emph{Computer Networks}, vol.~30, no. 1--7, pp. 107--117, 1998.

\bibitem{Langville2006-GPR}
A.~Langville and C.~Meyer, \emph{Google's PageRank and Beyond}.\hskip 1em plus
  0.5em minus 0.4em\relax Princeton University Press, 2006.

\bibitem{Manning2008-ITI}
C.~Manning, P.~Raghavan, and H.~Sch\"utze, \emph{Introduction to Information
  Retrieval}.\hskip 1em plus 0.5em minus 0.4em\relax Cambridge University
  Press, 2008.

\bibitem{Bauckhage2007-DFI}
C.~Bauckhage, ``{Distance-Free Image Retrieval Based on Stochastic Diffusion
  over Bipartite Graphs},'' in \emph{Proc. BMVC.}, 2007.

\bibitem{Jing2008-VAP}
Y.~Jing and S.~Baluja, ``{VisualRank: Applying PageRank to Large-scale Image
  Search},'' \emph{IEEE Trans. PAMI}, vol.~30, no.~11, pp. 1877--1890, 2008.

\bibitem{Bauckhage2008-PDC}
C.~Bauckhage, ``{Probabilistic Diffusion Classifiers for Object Detection},''
  in \emph{Proc. ICPR}.\hskip 1em plus 0.5em minus 0.4em\relax IEEE, 2008.

\bibitem{Neumann2011-MLS}
M.~Neumann, K.~Kersting, and B.~Ahmadi, ``{Markov Logic Sets: Towards Lifted
  Information Retrieval Using PageRank and Label Propagation},'' in \emph{Proc.
  AAAI}.\hskip 1em plus 0.5em minus 0.4em\relax AAAI Press, 2011.

\bibitem{Boldi2005-TRR}
P.~Boldi, ``{TotalRank: Ranking Without Damping},'' in \emph{Proc. WWW}.\hskip
  1em plus 0.5em minus 0.4em\relax ACM, 2005.

\bibitem{Bressan2010-CTD}
M.~Bressan and E.~Peserico, ``{Choosing the Damping, Choosing the Ranking?}''
  \emph{J. of Discrete Algorithms}, vol.~8, no.~2, pp. 199--213, 2010.

\bibitem{BaezaYates2006-GPR}
R.~Baeza-Yates, P.~Boldi, and C.~Castillo, ``{Generalizing PageRank: Damping
  Functions for Link-Based Ranking Algorithms},'' in \emph{Proc. SIGIR}.\hskip
  1em plus 0.5em minus 0.4em\relax ACM, 2006.

\bibitem{Gleich2014-ADS}
D.~Gleich and R.~Rossi, ``{A Dynamical System for PageRank with Time-Dependent
  Teleportation},'' \emph{Internet Mathematics}, vol.~10, no. 1--2, pp.
  188--217, 2014.

\end{thebibliography}

\end{document}